\documentclass[journal]{IEEEtran}
\usepackage{amsmath}
\usepackage{graphicx}
\usepackage{epsfig}
\usepackage{amssymb}
\usepackage{amsthm}
\usepackage{verbatim}
\usepackage{lettrine}
\usepackage{flushend,cuted}
\usepackage[top=1in, bottom=0.8in, left=0.5in, right=0.5in]{geometry}

\usepackage{url}
\usepackage{multirow}
\usepackage{cite}
\usepackage{mathrsfs}
\usepackage{cases}
\usepackage{caption}
\usepackage[english]{babel}
\usepackage[utf8x]{inputenc}
\usepackage{tikz}
\usepackage{nomencl}
\makenomenclature

\begin{document}

\title{Transient Stability Assessment Using Individual Machine Equal Area Criterion Part II: Stability Margin
}
{\author{Songyan Wang, Jilai Yu, Wei Zhang \emph{ Member, IEEE}

\thanks{

S. Wang is with Department of Electrical Engineering, Harbin Institute of Technology, Harbin 150001, China (e-mail: wangsongyan@163.com).

J. Yu is with Department of Electrical Engineering, Harbin Institute of Technology, Harbin 150001, China (e-mail: yupwrs@hit.edu.cn).

W. Zhang is with Department of Electrical Engineering, Harbin Institute of Technology, Harbin 150001, China (e-mail: wzps@hit.edu.cn).
}
}
\maketitle
\vspace{-15pt}
\begin{abstract}

In the second part of this two-paper series, the stability margin of a critical machine and that of the system are first proposed, and then the concept of non-global stability margin is illustrated. Based on the crucial statuses of the leading unstable critical machine and the most severely disturbed critical machine, the critical stability of the system from the perspective of an individual machine is analyzed. In the end of this paper, comparisons between the proposed method and classic global methods are demonstrated.

\end{abstract}


\nomenclature[1]{IMEF}{Individual machine energy function}
\nomenclature[2]{PEF}	{Partial energy function}
\nomenclature[3]{OMIB}	{One-machine-infinite-bus}
\nomenclature[4]{EAC}	{Equal area criterion}
\nomenclature[5]{COI}	{Center of inertia}
\nomenclature[6]{DLP}	{Dynamic liberation point}
\nomenclature[7]{DSP}	{Dynamic stationary point}
\nomenclature[8]{UEP}	{Unstable equilibrium point}
\nomenclature[9]{RUEP}	{Relevant UEP}
\nomenclature[10]{CUEP}	{Controlling UEP}
\nomenclature[11]{SVCS}	{Single-machine-and-virtual-COI-machine subsystem}
\nomenclature[12]{EEAC}	{Extended equal area criterion}
\nomenclature[13]{SIME}	{Single machine equivalence}
\nomenclature[14]{MOD}	{Mode of disturbance}
\nomenclature[15]{TSA}	{Transient stability assessment}
\nomenclature[16]{P.E.}	{Potential energy}
\nomenclature[17]{K.E.}	{Kinetic energy}
\nomenclature[18]{CCT}	{Critical fault clearing time}

\printnomenclature

\begin{center}
  \textsc{Abbreviation}
\end{center}

\begin{table}[!htbp]
\captionsetup{name=\textsc{Table}}
\normalsize
\vspace{5pt}
\setlength{\belowcaptionskip}{0pt}
\vspace{-4pt}
\setlength{\abovecaptionskip}{2pt}
\begin{tabular}{ll}
{COI}&{Center of inertia} \\

{CCT}&{Critical clearing time}\\

{CDSP}&{DSP of the critical stable machine}\\

{CUEP}&{Controlling UEP}\\

{DLP}&{Dynamic liberation point}\\

{DSP}&{Dynamic stationary point}\\

{EAC}&{Equal area criterion}\\

{IEEAC}&{Integrated extended EAC}\\

{IMEAC}&{Individual-machine EAC}\\

{IMEF}&{Individual machine energy function}\\

{IVCS}&{Individual machine-virtual COI machine system}\\

{LOSP}&{Loss-of-synchronism point}\\

{OMIB}&{One-machine-infinite-bus}\\

{PEF}&{Partial energy function}\\

{TSA}&{Transient stability assessment}\\

{UEP}&{Unstable equilibrium point}\\
\end{tabular}
\end{table}

\section{Introduction}

This is the second paper of the two-paper series dealing with power system transient stability by using IMEAC. In the first paper \cite{Wang2017Stability}, the mapping between IMEAC and the multi-machine system trajectory is established, and the unity principle of individual-machine stability and system stability is proposed.

The Ref. \cite{Stanton1989Analysis,Fouad1989Critical} was a milestone in the history of individual-machine methods because some crucial conjectures and hypothesis were first proposed in these two papers. Yet, a significant imperfection of the two papers is that Stanton did not explicitly explain the crucial transient stability concepts regarding the system in the sense of an individual machine, leaving these concepts missing in the two papers. These unsolved issues incurred lots of confusion and controversial problems when using individual-machine methods in TSA.

Following conclusions of the first paper, in this paper, first, the stability margin of a critical machine and that of the system are defined. The stability margin of the system is defined as a multi-dimensional vector that consists of stability margin of each critical machine in the system. Second, the important statuses of the most severely disturbed machine (MDM) and the leading unstable machine (LUM) are analyzed. According to unity principle, the non-global stability margin is defined and its application in TSA is also demonstrated. For the proposed method, the stability judgement of the system can be independent of the calculation of the stability margin of the system. And the proposed method allows system operators to neglect monitoring some critical machines when judging the stability of the system under certain circumstances. Third, the application of the proposed method in the computation of CCT is analyzed. We prove that the critical stability of the system is precisely identical to the critical stability of MDM. In the end of the paper comparisons of the proposed method with CUEP method and IEEAC method are demonstrated.

Contributions of this paper are summarized as follows:

(i) The stability margin of the system consists of multiple stability margins of critical machines of the system is first defined in this paper, and this definition of multi-dimensional vector enables the utilization of the non-global margin in TSA;

(ii) MDM and LUM are analyzed in this paper, and it is proved that they might be two different machines for some cases, which clarifies the historical misunderstanding in individual methods that MDM and LUM are the same machine;

(iii) According to unity principle, it is proved that the critical stability state of only one or a few MDMs that determines the critical stability of the system in this paper, and based on this, an approach to analyze the critical stability of the system using MDM is proposed.

The remaining paper is organized as follows. In Section II, the definitions of the stability margin of a critical machine and that of the system are provided. In Section III, the non-global stability margin is analyzed. In Section IV, general procedures of the proposed method are provided. In Section V, the application of the proposed method in TSA is demonstrated. In Section VI, the proposed method is utilized for CCT computation. In Sections VII and VIII, the proposed method is compared with the CUEP method and IEEAC method respectively. Conclusions and discussions are provided in Section IX.

\section{Stability Margin of the System}

\subsection{Stability margin of a critical machine}

\subsubsection{Stability margin of an unstable critical machine}

Using IMEAC, the stability margin of an unstable critical machine can be intuitively defined as:

\begin{equation}\label{Eq_Sta_Mar_USM}
\eta_{i}=(A_{DECi}-A_{ACCi})/A_{ACCi}
     \setlength{\abovedisplayskip}{1pt}
  \setlength{\belowdisplayskip}{1pt}
\end{equation}

The margin definition in (\ref{Eq_Sta_Mar_USM}) ensures that $\eta_{i}<0$ if a critical machine goes unstable as $A_{DECi}$ is smaller than $A_{ACCi}$ for an unstable critical machine \cite{Wang2017Stability}.

\subsubsection{Stability margin of a stable critical machine}

For a stable critical machine, the $A_{DECi}$ equals to $A_{ACCi}$ in the actual Kimbark curve of the machine \cite{Wang2017Stability}. Yet, the deceleration area of the machine may still possess a certain ``margin” after DSP occurs even though the acceleration energy during fault-on period is totally absorbed at DSP, as shown in Fig. \ref{Fig_Sta_Mar_SCM_1}.

\begin{figure}
\vspace{5pt}
\captionsetup{name=Fig.,font={small},singlelinecheck=off,justification=raggedright}
\includegraphics[width=3.5in,height=2.8in,keepaspectratio]{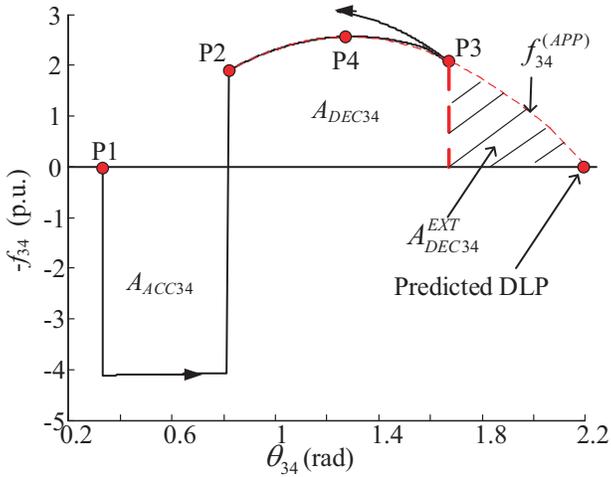}\\
  \setlength{\abovecaptionskip}{-5pt}
  \setlength{\belowcaptionskip}{0pt}
  \vspace{-2pt}
  \caption{  Stability margin of a stable critical machine (Machine 34, [TS-1, bus-34, 0.180s])}
  \label{Fig_Sta_Mar_SCM_1}
\end{figure}

For a stable critical machine, the Kimbark curve of the critical machine prior to DSP can be approximated by a quadratic function:

\begin{equation}\label{Eq_Kim_Cur_Qua}
f_{i}^{APP}=a_{i}\theta_{i}^{2}+b_{i}*\theta_{i}+c_{i}
     \setlength{\abovedisplayskip}{1pt}
  \setlength{\belowdisplayskip}{1pt}
\end{equation}

The parameters in (\ref{Eq_Kim_Cur_Qua}) can be identified via three points, i.e., fault clearing point (P2), DSP (P3) and the point with maximum $-f_{i}$ (P4). Further, the stability margin of a stable critical machine can be defined as:

\begin{equation}\label{Eq_Sta_Mar_CSM}
\eta_{i}=(A_{DECi}+A_{DECi}^{EXT}-A_{ACCi})/A_{ACCi}=A_{DECi}^{EXT}/A_{ACCi}
     \setlength{\abovedisplayskip}{1pt}
  \setlength{\belowdisplayskip}{1pt}
\end{equation}
\newline
where
\newline
$A_{DECi}^{EXT}=\int_{\theta_{i}^{DSP}}^{\theta_{i}^{DLP(PRED)}}[-f_{i}^{APP}]d\theta_{i}, A_{DECi}=A_{ACCi}$.

In (\ref{Eq_Sta_Mar_CSM}), the margin definition ensures that $\eta_{i}>0$ if a critical machine is stable, and $\eta_{i}=0$  if a critical machine is critical stable.

Consider that application of $f_{i}^{APP}$ may result in approximation error, in the following analysis the Kimbark curve of a stable critical machine prior to DSP still uses the original Kimbark curve and $f_{i}^{APP}$  is only used for the computation of $A_{DECi}^{EXT}$.

\subsection{Stability margin of the system}

Since the system operators focus on the swing stability of each critical machine in parallel in TSA, they may only target the IMEAC of each critical machine. Therefore, the stability margin of the system can be defined as a multi-dimensional vector that comprises of margins of all critical machines in the system:

\begin{equation}\label{Eq_Sta_Mar_Sys}
\boldsymbol{\eta}_{sys}=[\eta_{i}] \quad i\in{\Omega_{c}}
     \setlength{\abovedisplayskip}{1pt}
  \setlength{\belowdisplayskip}{1pt}
\end{equation}
\newline
where $\Omega_{c}$ is the set of all critical machines in the system.

Following unity principle the stability judgement of the system can be given as: all $\eta_{i}>0$  means the system is stable; one or a few $\eta_{i}=0$  with the rest of $\eta_{i}>0$ means the system is critical stable; one or more $\eta_{i}<0$  means the system goes unstable.

\subsection{Machine-by-machine margin calculation}

 From the analysis in the first paper, along post-fault system trajectory the DLPs and DSPs of critical machines occur machine-by-machine. Since the actual shape of the Kimbark curve of a machine in a complete swing could be formed only when the DSP or DLP occurs, accordingly $\eta_{i}$ of each critical machine should also be calculated in a ``machine-by-machine” way. An example of calculating the stability margin of the system during TSA is shown in Fig. \ref{Fig_MbM_Sta_Jud_2}.

\begin{figure}
\vspace{5pt}
\captionsetup{name=Fig.,font={small},singlelinecheck=off,justification=raggedright}
\includegraphics[width=3.5in,height=2.8in,keepaspectratio]{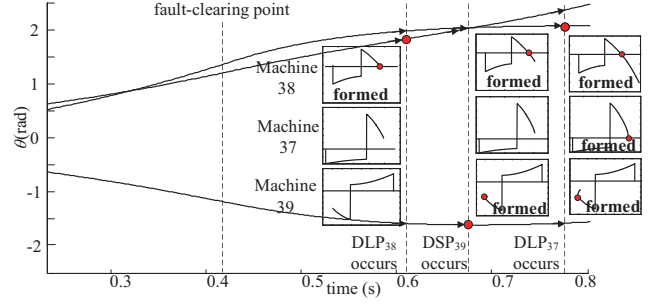}\\
  \setlength{\abovecaptionskip}{-5pt}
  \setlength{\belowcaptionskip}{0pt}
  \vspace{-2pt}
  \caption{ Machine-by-machine stability judgement along post-fault system trajectory [TS-1, bus-2, 0.43s]}
  \label{Fig_MbM_Sta_Jud_2}
\end{figure}

As shown in Fig. \ref{Fig_MbM_Sta_Jud_2}, after fault clearing the system operator only monitors Machines 37, 38 and 39 as they are critical machines. Along post-fault system trajectory the system operator may focus on following instants.

\emph{DLP38 occurs (0.614s)}: Machine 38 is judged as unstable, $\eta_{38}$ is obtained. Yet, $\boldsymbol{\eta}_{sys}$  is unknown due to the lack of $\eta_{37}$ and $\eta_{39}$.

\emph{DSP39 occurs (0.686s)}: Machine 39 is judged as stable, $\eta_{39}$  is obtained. $\boldsymbol{\eta}_{sys}$  is unknown due to the lack of $\eta_{37}$ .

\emph{DLP37 occurs (0.777s)}: Machine 37 is judged as unstable, $\eta_{37}$  is obtained. $\boldsymbol{\eta}_{sys}$ is obtained.

The computation of the margin of each critical machine is shown in Table \ref{Tab_Sta_Mar_CM_1}.

\begin{table}
\captionsetup{name=\textsc{Table},font={small}}
\vspace{5pt}
\centering
\setlength{\belowcaptionskip}{0pt}
\caption{\textsc{ Stability margin of critical machines}}
\vspace{-4pt}
\setlength{\abovecaptionskip}{2pt}
\begin{tabular}{ccccc}
\hline
Machine No.  & $A_{ACCi} (p.u.)$ & $A_{DECi} (p.u.)$ & $A_{ACCi}^{EXT} (p.u.)$ & $\eta_{i}$ \\
\hline
 38       & 1.33            & 0.54  & N/A & -0.594          \\
37	&2.09	&2.08	&N/A	&-0.005 \\
39	&3.22	&3.22	&1.57	&0.489\\
\hline
\end{tabular}
\label{Tab_Sta_Mar_CM_1}
\end{table}

Based on the margins of all critical machines in the system showing in Table \ref{Tab_Sta_Mar_CM_1}, $\boldsymbol{\eta}_{sys}$  is finally computed as [-0.594, -0.005, 0.489].

As for online security control, the controlling objective is to ensure that the entire system maintains stable, i.e., all critical machines in the system should be maintained stable (the controlling action should be deployed immediately once an unstable critical machine is observed). Thereby, the security control based on IMEAC method strongly relies on the computation of $\boldsymbol{\eta}_{sys}$. However, under certain circumstances it is possible that the system operator is only interested in the stability state of only one or a few most severely disturbed critical machines rather than all critical machines, as will be discussed in the following section.

\section{Non-global stability margin}

\subsection{Status of a Critical Machine }

 During post-fault transient period, two specific critical machines are crucial, which can partially represent the transient stability characteristics of all critical machines in the system. The first one is the most severely disturbed machine (MDM), with the second one being leading unstable machine (LUM).

 \subsubsection{MDM}

\begin{equation}\label{Eq_MDM}
{\eta}_{MDM}=min[\eta_{i}\mid i\in{\Omega_{c}}]
     \setlength{\abovedisplayskip}{1pt}
  \setlength{\belowdisplayskip}{1pt}
\end{equation}

  \subsubsection{LUM}

  The LUM is the unstable critical machine whose DLP occurs first among all unstable critical machines:
  \begin{equation}\label{Eq_LUM}
t_{DLP_{LUM}}=min[t_{DLP_{i}}\mid i\in{\Omega_{c}}]
     \setlength{\abovedisplayskip}{1pt}
  \setlength{\belowdisplayskip}{1pt}
\end{equation}

The reasons why MDM and LUM are important for transient stability analysis are as follows.

\emph{(i) The stability state of MDM is identical to the stability state of the system.}

Following unity principle, the system can be judged as stable if MDM is stable, the system can be judged as critical stable if MDM is critical stable, and the system can be judged as unstable if MDM goes unstable.

\emph{(ii) The LOSP of the system is identical to the DLP of LUM.}

The LUM is the first-going-unstable critical machine that separates from the system. Once LUM appears, the system operator can ascertain that the system goes unstable, and DLP of the LUM can be thence defined as the leading LOSP.

From analysis above, the two crucial transient stability characteristics of a multi-machine system, i.e. the stability state of the system and leading LOSP, respectively, are fully embodied in MDM and LUM.

For most unstable cases, the MDM and LUM may be the same machine because the residual K.E. at DLP of MDM would be quite high, which is most likely to drive it to go unstable first along time horizon (also is LUM). Yet, for some cases it is still possible that MDM and LUM might be two different machines.
\subsection{Non-global Stability Margin  }

\subsubsection{ Non-global stability margin for TSA (the system operator misses monitoring some critical machines)}

If the critical machines in the system are not all monitored, the stability margin of the system could only be depicted in an incomplete non-global form:

\begin{equation}\label{Eq_MDM}
\boldsymbol{\eta}_{sys}=[\eta_{i}] \quad i\in{\Omega_{non}} \quad \Omega_{non}\subset{\Omega_{c}}]
     \setlength{\abovedisplayskip}{1pt}
  \setlength{\belowdisplayskip}{1pt}
\end{equation}
\newline
where $\Omega_{non}$ is the set of non-globally monitored critical machines.

In (7), $\boldsymbol{\eta}_{non}$ is defined such that only parts of critical machines are monitored. Under this not-all-critical-machines monitoring circumstances, $\boldsymbol{\eta}_{sys}$  cannot be evaluated since it is defined by all critical machines. However, following unity principle, it is still possible that the stability state of the system can be judged as long as any one unstable critical machine is included in $\boldsymbol{\eta}_{non}$.

The analysis above is crucial because it reveals that: for the proposed method, the unity principle allows \emph{stability judgement of the system to be independent of the calculation of the stability margin of the system.
}

A demonstration of the not-all-critical-machines monitoring is shown in Fig. \ref{Fig_Dem_NCM_Mon_3}.

\begin{figure}
\vspace{5pt}
\captionsetup{name=Fig.,font={small},singlelinecheck=off,justification=raggedright}
\includegraphics[width=3.5in,height=2.8in,keepaspectratio]{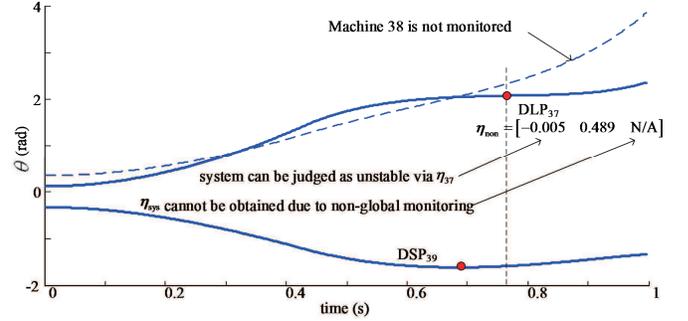}\\
  \setlength{\abovecaptionskip}{-5pt}
  \setlength{\belowcaptionskip}{0pt}
  \vspace{-2pt}
  \caption{ Demonstration of not-all-critical-machines monitoring [TS-1, bus-2, 0.43s]}
  \label{Fig_Dem_NCM_Mon_3}
\end{figure}

In this case Machine 38 is not monitored by the system operator, thus $\boldsymbol{\eta}_{sys}$  cannot be obtained. Yet, the system still can be judged as unstable once $\text{DLP}_37$ occurs.

In fact, judging the stability of the system by using only one unstable critical machine and \emph{neglecting} the rest of critical machines is a kind of ``transient information missing”. However, once MDM and LUM are monitored and both machines are included in $\Omega_{non}$, the system operator is still able to grasp the key transient information of the system, i.e., the stability state and leading LOSP of the system. Tolerating transient information missing can be seen as a distinctive characteristic of the proposed method, making it quite robust in TSA.

\subsubsection{Non-global stability margin for CCT computation (the system operator focuses on MDM on purpose)}

Following the definition of MDM, the stability state of MDM is identical to the stability state of the system. From the perspective of individual-machine analysts, the main role of MDM is its application in CCT computation, because \emph{the critical stability of the system is fully determined by the critical stability of MDM, according to unity principle.} Therefore, during iterations of fault clearing time when computing CCT, the system operator may only target MDM without observing the rest of critical machines, and the non-global stability margin for the CCT computation can be directly depicted as:
\begin{equation}\label{Eq_Sta_NON}
\boldsymbol{\eta}_{non}=\eta_{MDM}
     \setlength{\abovedisplayskip}{1pt}
  \setlength{\belowdisplayskip}{1pt}
\end{equation}

In (\ref{Eq_Sta_NON}), the stability of the whole system is fully represented by that of the MDM when computing CCT of the system. Detailed analysis about CCT computation is provided in Section VI.

\section{Procedures of the proposed method in transient stability analysis}

\subsection{Application of the proposed method in TSA}
For a certain occurred fault in OTSA, the procedures of parallel monitoring using the proposed method are outlined as follows.

\emph{Step 1:}	Monitor all critical machines in parallel after fault clearing.

\emph{Step 2:}	Along post-fault system trajectory the stability of all critical machines in the system is judged in a machine-by-machine way. The stability of the machine is identified via the occurrence of DLP or DSP, and $\eta_i$ of the critical machine is calculated via IMEAC.

\emph{Step 3:}	If one or more DLPs occur, the first occurred DLP is defined as the leading LOSP, and system is immediately judged as unstable. Meanwhile, the machine with the first occurred DLP is defined as LUM.

\emph{Step 4:}	If DLP does not occur and, instead, DSPs occur one after another until the Kimbark curve of the last critical machine is formed, the system can be judged as stable when the last DSP occurs.

\emph{Step 5:}	At the instant that the Kimbark curve of the last critical machine is formed, the MDM is identified and $\boldsymbol{\eta}_{sys}$ is finally obtained.

The procedures of not-all-critical-machines monitoring are almost the same with that of parallel monitoring. The only difference is that the critical machines are not all monitored for the not-all-critical-machines monitoring case.

\subsection{Application of the proposed method in CCT computation}
The procedures of the CCT computation using the proposed method are outlined as follows.

\emph{Step 1:}	For the first few iterations with the system being stable, the MDM is identified by finding the minimum $\eta_i$ among all stable critical machines in the system.

\emph{Step 2:}	Once MDM is identified, only MDM is monitored for the followed iterations.

\emph{Step 3:}	If $\eta_{MDM}>0$ when fault clearing time is $t_u$ and $\eta_{MDM}<0$ when fault clearing time is $t_u+\Delta{t}$, then $t_u$ is set as the CCT of the system.

\section{Simulations of the Proposed Method in TSA}

\subsection{ Parallel Monitoring }

 Here, the small-scale test network TS-1 is provided first to systematically demonstrate the application of the proposed method in OTSA. The fault is [TS-1, bus-21, 0.370s]. The system trajectory is shown in Fig. \ref{Fig_Sys_Tra_4}.

\begin{figure}
\vspace{5pt}
\captionsetup{name=Fig.,font={small},singlelinecheck=off,justification=raggedright}
\includegraphics[width=3.5in,height=2.8in,keepaspectratio]{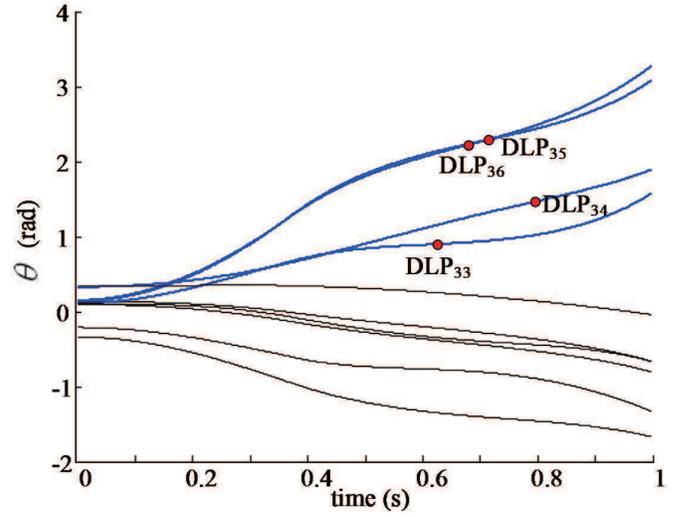}\\
  \setlength{\abovecaptionskip}{-5pt}
  \setlength{\belowcaptionskip}{0pt}
  \vspace{-2pt}
  \caption{  System trajectory [TS-1, bus-21, 0.37s]}
  \label{Fig_Sys_Tra_4}
\end{figure}

After fault clearing, Machines 33-36 are defined as critical machines. Using the proposed method, along time horizon the system operators focus on following instants.

\emph{DLP33 occurs (0.611s)}: (i) Machine 33 is judged as unstable, and $\eta_{33}$ is obtained. (ii) Machine 33 is identified as LUM.

\emph{DLP36—DLP35 occur (0.671s—0.710s)}: Corresponding critical machines are judged as unstable, and $\eta_{i}$  of them are obtained one after another.

\emph{DLP34 occurs (0.795s)}: (i) Machine 34 is judged as unstable, and $\eta_{34}$  is obtained. (ii) Machine 34 is identified as MDM. (iii) $\boldsymbol{\eta}_{sys}$ is obtained.

The stability of the system along time horizon is judged as below:

\emph{DLP33 occurs (0.611s)}: DLP33 is identified as the leading LOSP, and system is judged as unstable.

\emph{DLP34 occurs (0.795s)}: $\boldsymbol{\eta}_{sys}$  is obtained.

Kimbark curves of critical machines are shown in Figs. \ref{Fig_Sim_Kim_Cur_5} (a)-(d), respectively. The stability margins of critical machines are shown in Table \ref{Tab_Sta_Mar_CM_2}.

\begin{figure}
\captionsetup{name=Fig.,font={small},singlelinecheck=off,justification=raggedright}
  \includegraphics[width=3.5in,height=5.4in,keepaspectratio]{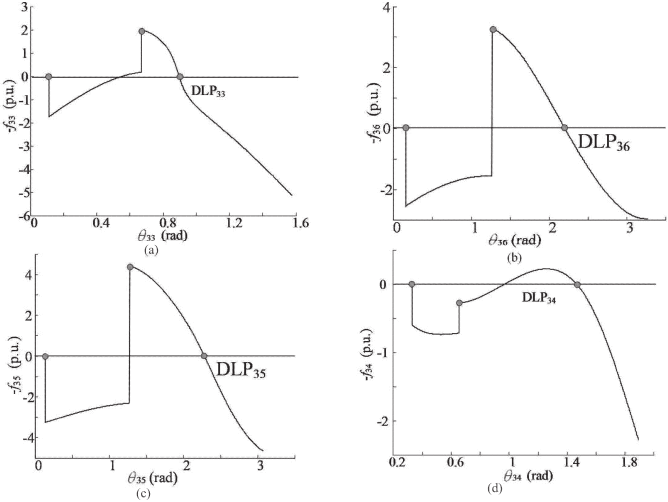}\\
  \setlength{\abovecaptionskip}{-5pt}
  \setlength{\belowcaptionskip}{0pt}
  \vspace{-2pt}
  \caption{  Simulated Kimbark curves [TS-1, bus-21, 0.370s]. (a-d) Kimbark curves of Machines 33, 36, 35 and 34.}
  \label{Fig_Sim_Kim_Cur_5}
\end{figure}

\begin{table}
\captionsetup{name=\textsc{Table},font={small}}
\vspace{5pt}
\centering
\setlength{\belowcaptionskip}{0pt}
\caption{\textsc{Stability margin of critical machines}}
\vspace{-4pt}
\setlength{\abovecaptionskip}{2pt}
\begin{tabular}{cccc}
\hline
Machine No.  & $A_{ACCi} (p.u.)$ & $A_{DECi} (p.u.)$  & $\eta_{i}$ \\
\hline

 33	&0.34	&0.33	&-0.03       \\
36	&2.09	&1.83	&-0.12 \\
35	&3.03	&2.82	&-0.07 \\
34	&0.2816	&0.0717	&-0.75 \\
\hline
\end{tabular}
\label{Tab_Sta_Mar_CM_2}
\end{table}

As shown in Table \ref{Tab_Sta_Mar_CM_2}, at the moment of DLP34 occurs, $\boldsymbol{\eta}_{sys}$  is finally calculated as [-0.03, -0.12, -0.07, -0.75].

\subsection{Discussion of MDM and LUM}

For the case shown in Fig. \ref{Fig_Sys_Tra_4}, Machines 34 and 33 are identified as MDM and LUM, respectively, The occurrence of MDM and LUM along time horizon is shown in Fig. \ref{Fig_MDM_LUM_Ide_6}.

\begin{figure}
\vspace{5pt}
\captionsetup{name=Fig.,font={small},singlelinecheck=off,justification=raggedright}
  \includegraphics[width=3.5in,height=5.4in,keepaspectratio]{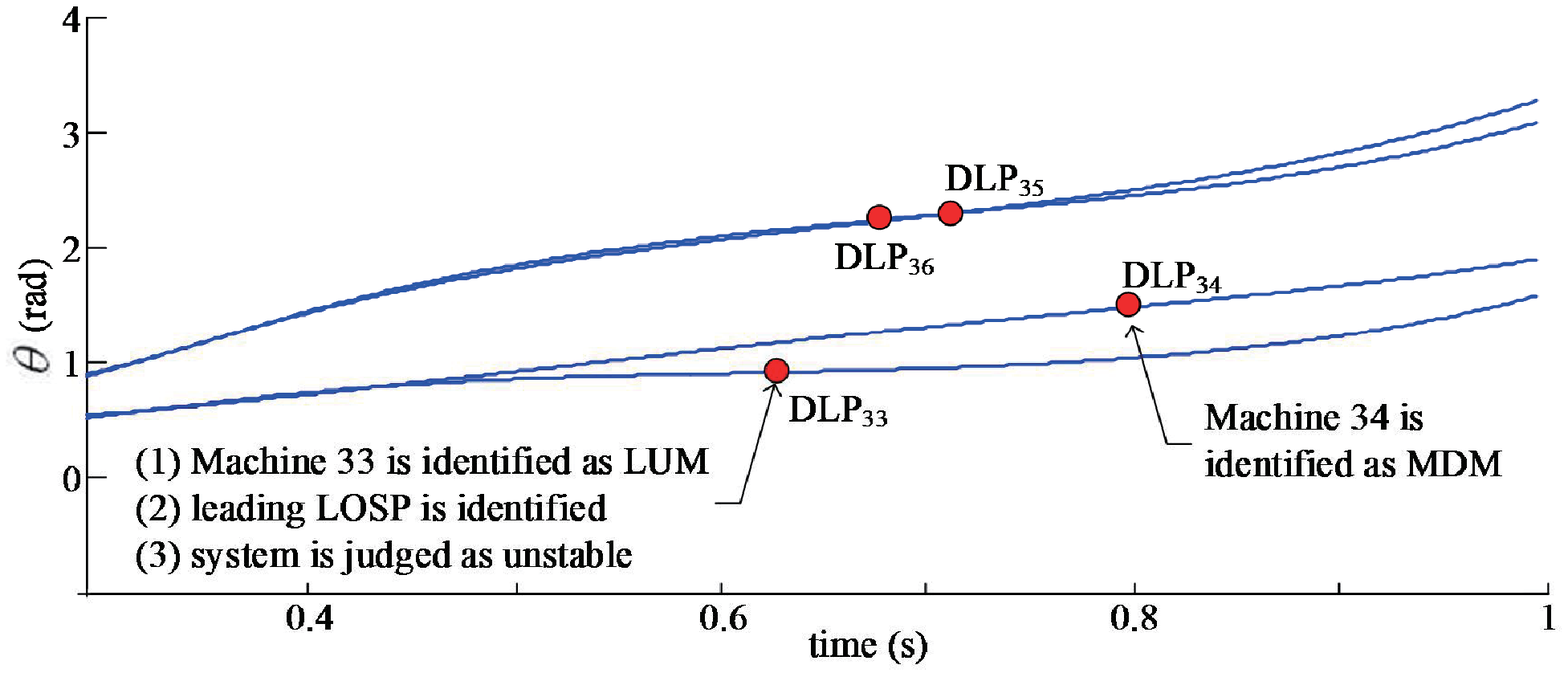}\\
  \setlength{\abovecaptionskip}{-5pt}
  \setlength{\belowcaptionskip}{0pt}
  \vspace{-2pt}
  \caption{Identification of MDM and LUM [TS-1, bus-21, 0.37s]}
  \label{Fig_MDM_LUM_Ide_6}
\end{figure}

The result shown in Fig. \ref{Fig_MDM_LUM_Ide_6} validates that the MDM and LUM might not be the same machine under certain circumstance. Especially, for the case in Fig. \ref{Fig_Sys_Tra_4}, Machine 33 possesses the highest margin but it is the first one to go unstable, whereas Machine 34 possesses the lowest margin but it is the last one to go unstable. As shown in Fig. \ref{Fig_MDM_LUM_Ide_6}, Machine 34 is still decelerating at the instant of $\text{DLP}_{33}$ (0.611s), which indicates that there might be little correlation between MDM and LUM in some simulation cases.

From simulation results above one can see, along actual post-fault system trajectory the MDM can be identified only when the stability margin of the last critical machine is calculated, because the determination of the minimum margin should be based on acquisition of the stability margin of all critical machines. Comparatively, the identification of LUM may be more ``important” than MDM in TSA because both the instability of the system and leading LOSP can be immediately identified when LUM occurs without waiting for the occurrence of MDM. And the system operators should already be prepared to take proactive controlling actions at the instant when the leading LOSP occurs.

\subsection{Not-all-critical-machines Monitoring }

For the case in Fig. \ref{Fig_Sys_Tra_4}, assume the system operator neglects monitoring Machine 34 after fault clearing, then only three critical machines are monitored. Under this circumstance, the instability of the system still can still be rightly identified when DLP33 occurs. Later, Machines 36 and 35 are both judged as unstable consecutively. Meanwhile, at the instant of DLP35 $\boldsymbol{\eta}_{non}$  is finally calculated as [-0.03, -0.12, -0.07, N/A].

From analysis above, although $\boldsymbol{\eta}_{sys}$  cannot be obtained due to the unmonitored Machine 34, both the stability state and leading LOSP of the system can still be rightly identified via the three monitored machines, which indicates that $\boldsymbol{\eta}_{non}$  could partially represent the margin of the system.

\section{Applications of the Proposed Method in CCT Computation}

In this section, a simulation case is provided to illustrate the relationship between the critical stability of the critical machines and that of the system. The CCT of the system for the fault [TS-1, bus-19] is 0.215s according to the time-domain simulation. Machines 33, 34 and 39 are critical machines. The critical stable and critical unstable system trajectory are shown in Figs. \ref{Fig_Sim_Tra_7} (a) and (b), respectively. Kimbark curves of critical machine 34 in both cases are shown in Figs. \ref{Fig_Kim_Dia_8} (a) and (b), respectively.

\begin{figure}
\vspace{5pt}
\captionsetup{name=Fig.,font={small},singlelinecheck=off,justification=raggedright}
  \includegraphics[width=3.5in,height=5.4in,keepaspectratio]{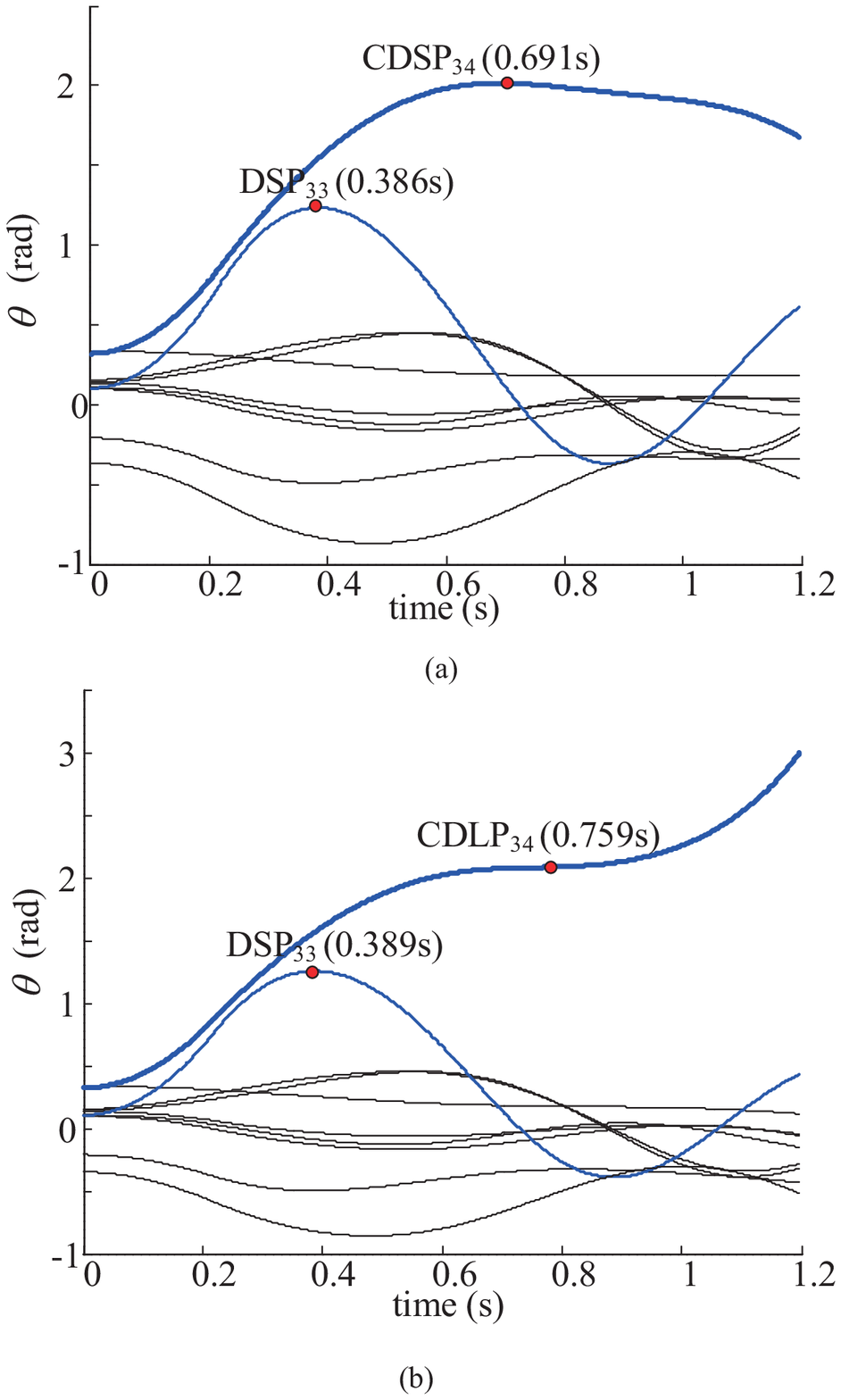}\\
  \setlength{\abovecaptionskip}{-5pt}
  \setlength{\belowcaptionskip}{0pt}
  \vspace{-2pt}
  \caption{Simulated system trajectory. (a) critical stable case. (b) critical unstable case}
  \label{Fig_Sim_Tra_7}
\end{figure}

\begin{figure}
\vspace{5pt}
\captionsetup{name=Fig.,font={small},singlelinecheck=off,justification=raggedright}
  \includegraphics[width=3.5in,height=2.4in,keepaspectratio]{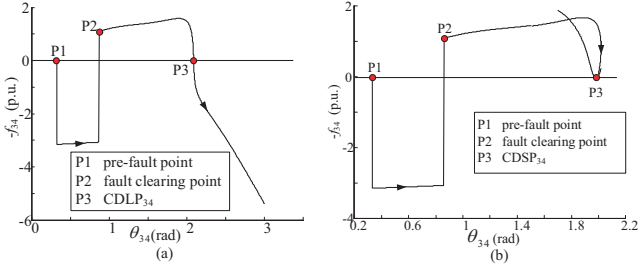}\\
  \setlength{\abovecaptionskip}{-5pt}
  \setlength{\belowcaptionskip}{0pt}
  \vspace{-2pt}
  \caption{Kimbark curve of stable couple machines}
  \label{Fig_Kim_Dia_8}
\end{figure}

In Figs. \ref{Fig_Sim_Tra_7} (a) and (b), we observe the critical stable and critical unstable system trajectories from the perspective of individual machine. When fault clearing time is 0.215s, Machine 34 (i.e., IMT34) is critical stable whereas Machines 33 and 39 are marginal stable, thus the system trajectory is kept stable, as shown in Fig. \ref{Fig_Kim_Dia_8} (a). When fault clearing time is slightly increased to 0.216s, Machines 33 and 39 are still marginal stable. Yet, Machine 34 starts separating from the system at $\text{CDLP}_{34}$ and finally goes unstable, causing the system to go unstable, as shown in Fig. \ref{Fig_Kim_Dia_8} (b). From analysis above, the critical stability of Machine 34 fully determines the critical stability of the system, and the real inflection point of the critical stable system trajectory is $\text{CDLP}_{34}$.

The variation of the $\text{IMT}_{34}$ with the change of fault clearing time is shown in Fig. \ref{Fig_Var_IMT_9}. The stability margins of critical machines are shown in Table \ref{Tab_Var_Sta_Mar_3}.

\begin{figure}
\vspace{5pt}
\captionsetup{name=Fig.,font={small},singlelinecheck=off,justification=raggedright}
  \includegraphics[width=3.5in,height=2.4in,keepaspectratio]{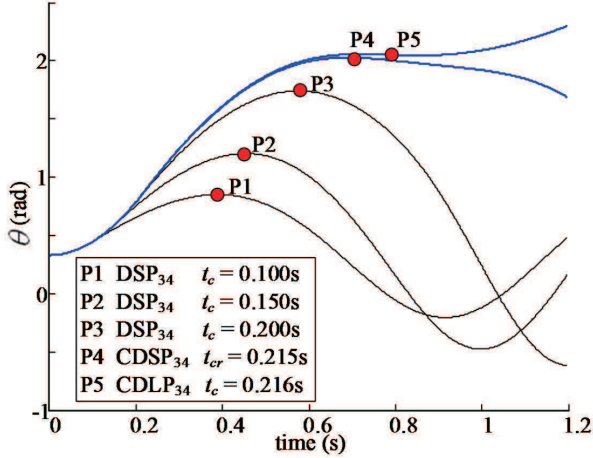}\\
  \setlength{\abovecaptionskip}{-5pt}
  \setlength{\belowcaptionskip}{0pt}
  \vspace{-2pt}
  \caption{Kimbark curve of the critical stable couple machines}
  \label{Fig_Var_IMT_9}
\end{figure}

\begin{table}
\captionsetup{name=\textsc{Table},font={small}}
\vspace{5pt}
\centering
\setlength{\belowcaptionskip}{0pt}
\caption{\textsc{Variations of the stability margins of critical machines with the increase of fault clearing time}}
\vspace{-4pt}
\setlength{\abovecaptionskip}{2pt}
\begin{tabular}{cccc}
\hline
CT(s)  & Machine No. & $\eta_{i}$  & MDM No. \\
\hline
\multirow{3}{*} {0.213}  & 34   & \underline{0.112 }& \multirow{3}{*} {34} \\
                         &33	&1.055 & \\
                         &39	&1.440 &  \\
\hline
\multirow{3}{*} {0.214}  & 34   &  \underline{0.104} & \multirow{3}{*} {34} \\
                         &33	&1.047 & \\
                         &39	&1.428 &  \\
\hline
\multirow{3}{*} {0.215}  & 34   &  \underline{0.068} & \multirow{3}{*} {34} \\
                         &33	&1.038 & \\
                         &39	&1.416 &  \\
\hline
\multirow{3}{*} {0.216}  & 34   &  \underline{-0.001} & \multirow{3}{*} {34} \\
                         &33	&1.026 & \\
                         &39	&1.405 &  \\
\hline
\multirow{3}{*} {0.216}  & 34   &  \underline{-0.005} & \multirow{3}{*} {34} \\
                         &33	&1.021 & \\
                         &39	&1.392 &  \\
\hline
\end{tabular}
\label{Tab_Var_Sta_Mar_3}
\end{table}

Following the procedures of the CCT computation as in Section IV.B, after first few iterations, it can be confirmed Machine 34 is the MDM as ${\eta}_{34}$  is much lower than ${\eta}_{33}$  and ${\eta}_{39}$  as shown in Table \ref{Tab_Var_Sta_Mar_3}, thus the system operator may only monitor Machine 34 in the followed iterations (${\eta}_{33}$  and ${\eta}_{39}$  are also shown in the table to demonstrate the variation of the margin). Later, with the increase of fault clearing time, Machine 34 is still kept as MDM all along. When fault clearing time is 0.216s, ${\eta}_{34}$  changes from positive to negative and Machine 34 becomes critical unstable, which causes the system to go critical unstable. Under this circumstance, Machine 34 is not only MDM but also LUM. Meanwhile, both ${\eta}_{33}$  and ${\eta}_{39}$  keep decreasing with the increase of fault clearing time. This is because both critical machines are disturbed more severely with the increase of fault clearing time. Yet, compared with ${\eta}_{34}$ , ${\eta}_{33}$  and ${\eta}_{39}$  are still quite high, which means that stability of Machines 33 and 39 will not affect the critical stability of the system. This validates that it is only the critical stability of Machine 34 that determines the critical stability of the system.

For some cases, it is possible that some critical machines might be highly correlated and they may all change from being critical stable to being critical unstable, as shown in Fig. \ref{Fig_Sim_Tra_10}.

\begin{figure}
\vspace{5pt}
\captionsetup{name=Fig.,font={small},singlelinecheck=off,justification=raggedright}
  \includegraphics[width=3.5in,height=4.8in,keepaspectratio]{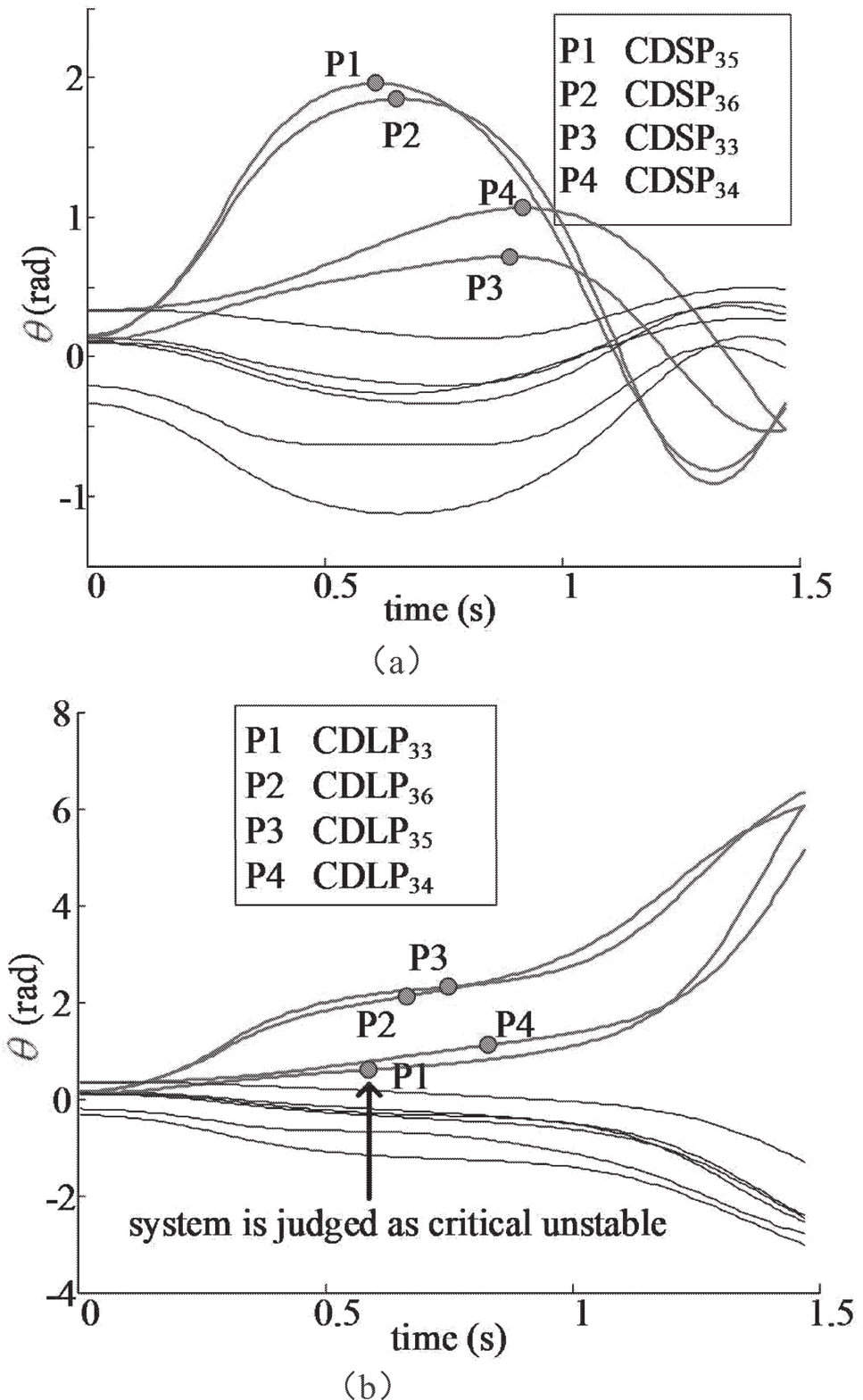}\\
  \setlength{\abovecaptionskip}{-5pt}
  \setlength{\belowcaptionskip}{0pt}
  \vspace{-2pt}
  \caption{  Simulated system trajectory [TS-1, bus-22, 0.295s], [TS-1, bus-22, 0.296s]. (a) critical stable case. (b) critical unstable case.}
  \label{Fig_Sim_Tra_10}
\end{figure}

For the case in Fig. \ref{Fig_Sim_Tra_10}, Machines 33-36 should all be determined as MDMs and thus should be monitored in parallel. However, the critical instability of the system can be identified once the LUM among these MDMs occurs without waiting for the occurrence of DLPs of other unstable critical machines. For instance, in this case Machines 33-36 all go critical unstable when fault clearing time is 0.30s, yet the critical instability of the system can be immediately identified when the DLP of the LUM (CDLP33) occurs at 0.58s, as shown in Fig. \ref{Fig_Sim_Tra_7} (b).

Detailed calculations of the CCT in different test systems are shown in Table \ref{Tab_Cpt_CCT_4}. The simulation step-size is set as 0.01s. Results reveal that the computed CCT is precisely identical to time domain simulations.

\begin{table}
\captionsetup{name=\textsc{Table},font={small}}
\vspace{5pt}
\centering
\setlength{\belowcaptionskip}{0pt}
\caption{\textsc{Calculations of CCT using proposed method}}
\vspace{-4pt}
\setlength{\abovecaptionskip}{2pt}
\begin{tabular}{ccccc}
\hline
Fault & Simulation & Proposed & MDM & LUM \\
location & ( s) &  method (s) & No. & No.\\
\hline
*TS-1, bus-34	&0.20	&0.20	&34	&34\\
*TS-1, bus-35	&0.30	&0.30	&35, 36	&36\\
*TS-1, bus-36	&0.26	&0.26	&36	&36\\
*TS-1, bus-37	&0.24	&0.24	&37	&37\\
TS-1, bus-4	    &0.44	&0.44	&31, 32	&32\\
TS-1, bus-15	&0.49	&0.49	&33-36	&33\\
TS-1, bus-21	&0.35	&0.35	&33-36	&33\\
TS-1, bus-24	&0.37	&0.37	&33-36	&33\\
*TS-2, bus-10	&0.19	&0.19	&2	&2\\
*TS-2, bus-12	&0.50	&0.50	&2	&2\\
*TS-2, bus-25	&0.34	&0.34	&4, 5	&5\\
*TS-2, bus-40	&0.57	&0.57	&8	&8\\
*TS-2, bus-66	&0.34	&0.34	&14	&14\\
TS-2, bus-5	    &0.27	&0.27	&2	&2\\
TS-2, bus-64	&0.71	&0.71	&12	&12\\
TS-2, bus-104	&0.37	&0.37	&20	&20\\
\hline
\multicolumn{5}{c}{(‘*’: Fault occurs at the terminal of the machine;``MDM No.”: \quad the \quad }\\
\multicolumn{5}{c}{MDM(s) when system is critical stable; ``LUM No”: the LUM when \qquad }\\
\multicolumn{5}{c}{system goes critical unstable) \qquad \qquad \qquad \qquad \qquad  \qquad  \qquad  \qquad  \qquad}\\
\end{tabular}
\label{Tab_Cpt_CCT_4}
\end{table}

\section{Comparison between CUEP Method and Proposed Method}

\subsection{CUEP Method}

 The analysis in this section is fully based on the classic simulation cases in Ref. \cite{Fouad1989Critical}. All parameters of CUEP method follow the forms in Ref. \cite{Fouad1989Critical}. The critical unstable case is set as [TS-2, bus-12, 0.510s] \cite{Fouad1989Critical}. Machines 2 and 3 are critical machines with only Machine 2 going critical unstable in this case.

In CUEP method, Fouad assumed that only critical machines might go unstable and the concept of ``approximate $\theta^{u}$” is proposed to initiate CUEP. To be specific, for the case above Fouad assumed that only Machines 2 and 3 might go unstable. In this way the possible combinations of the MODs are only three types, as shown in Fig. \ref{Fig_Pos_Mod_CUEP_11}.

\begin{figure*}
\captionsetup{name=Fig.,font={small},singlelinecheck=off,justification=raggedright}
  \includegraphics[width=6.98in,height=2.82in,keepaspectratio]{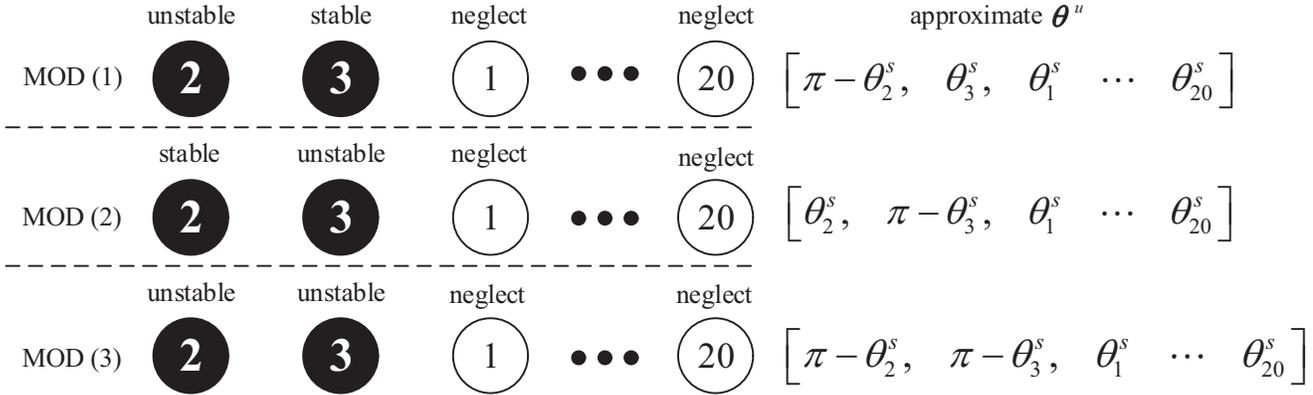}\\
    \setlength{\abovecaptionskip}{-5pt}
  \setlength{\belowcaptionskip}{0pt}
  \vspace{-5pt}
  \caption{Possible MODs in CUEP method}
  \label{Fig_Pos_Mod_CUEP_11}
\end{figure*}

As shown in Fig. \ref{Fig_Pos_Mod_CUEP_11}, the approximate $\theta^{u}s$ for initiation are $\underline{\theta_{2}^{u}}$, $\underline{\theta_{3}^{u}}$  and $\underline{\theta_{2,3}^{u}}$ \cite{Fouad1989Critical}. Further, the CUEP and the corresponding real MOD are identified via the computation of the lowest global critical energy. In this case the CUEP is computed as $\theta_{2}$, $\theta_{2,CUEP}=2.183$ rad and the real MOD is finally identified as only Machine 2 going unstable \cite{Fouad1989Critical}.

\subsection{Proposed Method }

Compared with the CUEP method that generates possible MODs first and then identifies the real MOD via the calculation of CUEP, the proposed method works in a more intuitive way, because the proposed method directly monitors the IMTs of Machines 2 and 3 in parallel, as shown in Fig. \ref{Fig_IMT_Mac_12}. Simulation results show that $\text{DSP}_{3}$ and $\text{DLP}_{2}$ occur at 0.613s and 0.948s, respectively. The Kimbark curves of the Machines 2 and 3 are shown in Figs. \ref{Fig_Sim_Kim_Cur_13} (a) and (b), respectively. The stability margins of Machines 2 and 3 are shown in Table \ref{Tab_Sta_Mar_CM_5}.

\begin{figure}
\captionsetup{name=Fig.,font={small},singlelinecheck=off,justification=raggedright}
  \includegraphics[width=3.7in,height=2.4in,keepaspectratio]{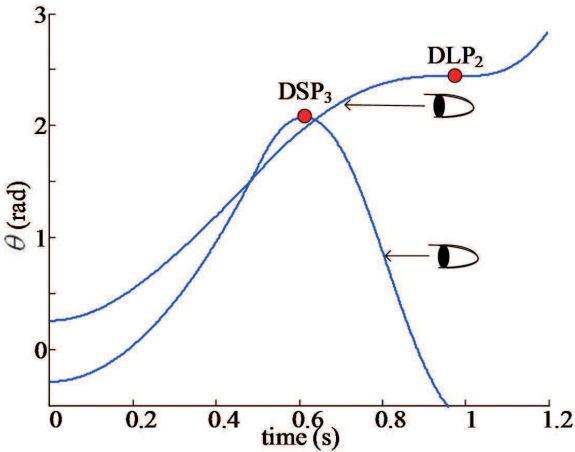}\\
  \setlength{\abovecaptionskip}{-5pt}
  \setlength{\belowcaptionskip}{0pt}
  \vspace{-5pt}
  \caption{IMTs of Machines 2 and 3}
  \label{Fig_IMT_Mac_12}
\end{figure}

\begin{figure}
\captionsetup{name=Fig.,font={small},singlelinecheck=off,justification=raggedright}
  \includegraphics[width=3.5in,height=2.4in,keepaspectratio]{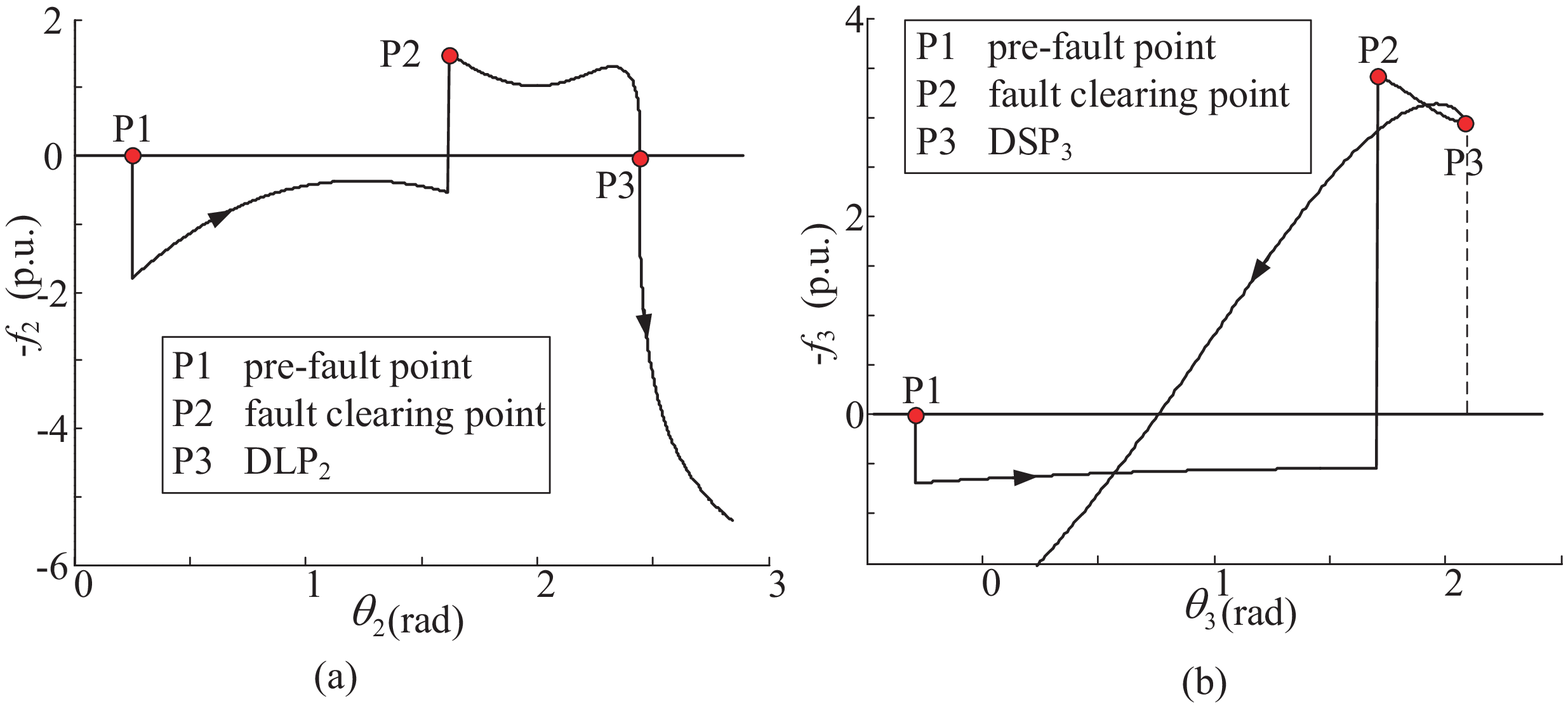}\\
  \setlength{\abovecaptionskip}{-5pt}
  \setlength{\belowcaptionskip}{0pt}
  \vspace{-5pt}
  \caption{ Simulated Kimbark curves (a) Machine 2. (b) Machine 3}
  \label{Fig_Sim_Kim_Cur_13}
\end{figure}

\begin{table}
\captionsetup{name=\textsc{Table},font={small}}
\vspace{5pt}
\centering
\setlength{\belowcaptionskip}{0pt}
\caption{\textsc{ Stability margin of critical machines}}
\vspace{-4pt}
\setlength{\abovecaptionskip}{2pt}
\begin{tabular}{ccccc}
\hline
Machine No.  & $A_{ACCi} (p.u.)$ & $A_{DECi} (p.u.)$ & $A_{ACCi}^{EXT} (p.u.)$ & $\eta_{i}$ \\
\hline
Machine 2	&0.97	&0.96	&N/A	&-0.01 \\
Machine 3	&1.17	&1.17	&1.03	&0.88\\
\hline
\end{tabular}
\label{Tab_Sta_Mar_CM_5}
\end{table}

From analysis above, Machine 3 is judged as stable at $\text{DSP}_{3}$. Later, Machine 2 is judged as unstable at $\text{DLP}_{2}$, which directly causes the system to go unstable. Thus the MOD is identified via the stability judgement of each critical machine without using CUEP.

\section{Comparison between IEEAC method and Proposed Method}

In this section, comparisons between IEEAC method and the proposed method are provided to demonstrate the application of the proposed method when identifying the inter-area instability in OTSA. The TS-3 is a practical 2766-bus, 146-unit interconnected system. SYSTEM\_LC is a regional system with 8 units, while SYSTM\_SD is a main system with 138 units. SYSTEM\_LC and SYSTM\_SD are connected through a 500 kV double-DC transmission lines. Five-order dynamic generator model with excitation and governor is utilized for simulation. The load consists of constant power load, constant impedance load, composite load and electric motor. Geographical layout of the interconnected system is shown in Fig. \ref{Fig_Geo_lay_Int_14}. The fault location is set at 90\% of Line LIAOC\_TANZ that is close to TANZ. Three phase short circuit event occurs at 0.00s and is cleared at 0.22s (CCT is 0.16s). In this case all machines in SYSTEM\_LC accelerate with respect to SYSTEM\_SD after fault is cleared, thus the separation mode of the interconnected system is a typical inter-area instability mode.

\begin{figure}
\vspace{5pt}
\captionsetup{name=Fig.,font={small},singlelinecheck=off,justification=raggedright}
  \includegraphics[width=3.6in,height=2.4in,keepaspectratio]{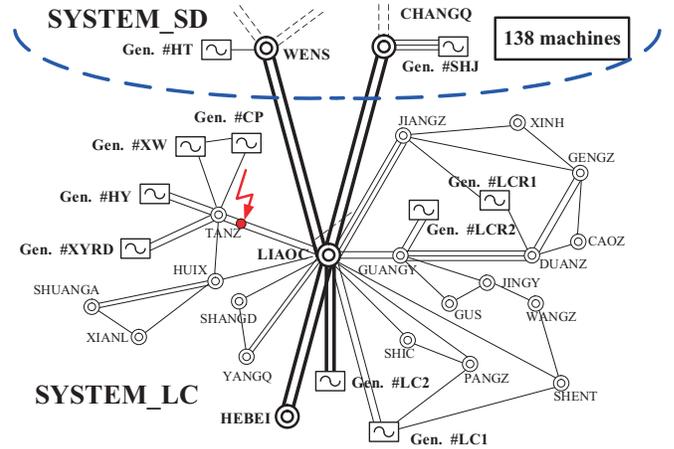}\\
  \setlength{\abovecaptionskip}{-5pt}
  \setlength{\belowcaptionskip}{0pt}
  \vspace{-5pt}
  \caption{Geographical layout of the interconnected system.}
  \label{Fig_Geo_lay_Int_14}
\end{figure}

\subsection{IEEAC Method}

The simulated rotor angles of the interconnected system in synchronous reference is shown in Fig. \ref{Fig_Rot_Ang_Int_15}.

\begin{figure}
\vspace{5pt}
\captionsetup{name=Fig.,font={small},singlelinecheck=off,justification=raggedright}
  \includegraphics[width=3.6in,height=2.4in,keepaspectratio]{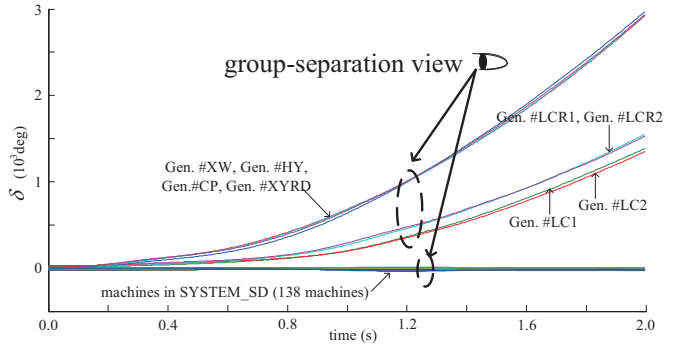}\\
  \setlength{\abovecaptionskip}{-5pt}
  \setlength{\belowcaptionskip}{0pt}
  \vspace{-5pt}
  \caption{Rotor angles of the interconnected system in synchronous reference [TS-3, line-LIAOC\_TANZ, 0.22s]}
  \label{Fig_Rot_Ang_Int_15}
\end{figure}

For simplification, $\Omega_{n_LC}$ is defined as the set with [Gen. \#XW, Gen. \#HY, Gen. \#CP, Gen. \# XYRD], $\Omega_{LC}$  is defined as the set with [Gen. \#LC1, Gen. \#LC2, Gen.\#LCR1, Gen. \#LCR2], and $\Omega_{SD}$  is defined as the set with all machines in SYSTEM\_SD. Using IEEAC method, all machines in the interconnected system after fault clearing are separated into the critical group $\Omega_{A}$  and the non-critical group$\Omega_{S}$ . Possible group separation modes are given as follows:

Mode-1: $\Omega_{A}$  is set as $\Omega_{LC}$, and $\Omega_{S}$ is set as $\Omega_{s_LC}\cup\Omega_{SD}$;
Mode-2: $\Omega_{A}$ is set as $\Omega_{LC}\cup\Omega_{s_LC}$, and $\Omega_{S}$ is set as $\Omega_{SD}$.

After calculating the stability margin of the OMIB system for the above two modes \cite{Yin2011An}, the Mode-2 whose OMIB system possesses a lower margin is finally identified as the dominate group separation mode. In this mode, eight machines in $\Omega_{LC}\cup\Omega_{s_LC}$ and 138 machines in $\Omega_{SD}$ are equated as Machine-A and Machine-S, respectively. The Kimbark curve of the equated OMIB system is shown in Fig. \ref{Fig_Kim_Cur_OMIB_16}. Notice that the equivalent Pm in the figure is not a horizontal line as the governors are deployed.

\begin{figure}
\vspace{5pt}
\captionsetup{name=Fig.,font={small},singlelinecheck=off,justification=raggedright}
  \includegraphics[width=3.6in,height=2.4in,keepaspectratio]{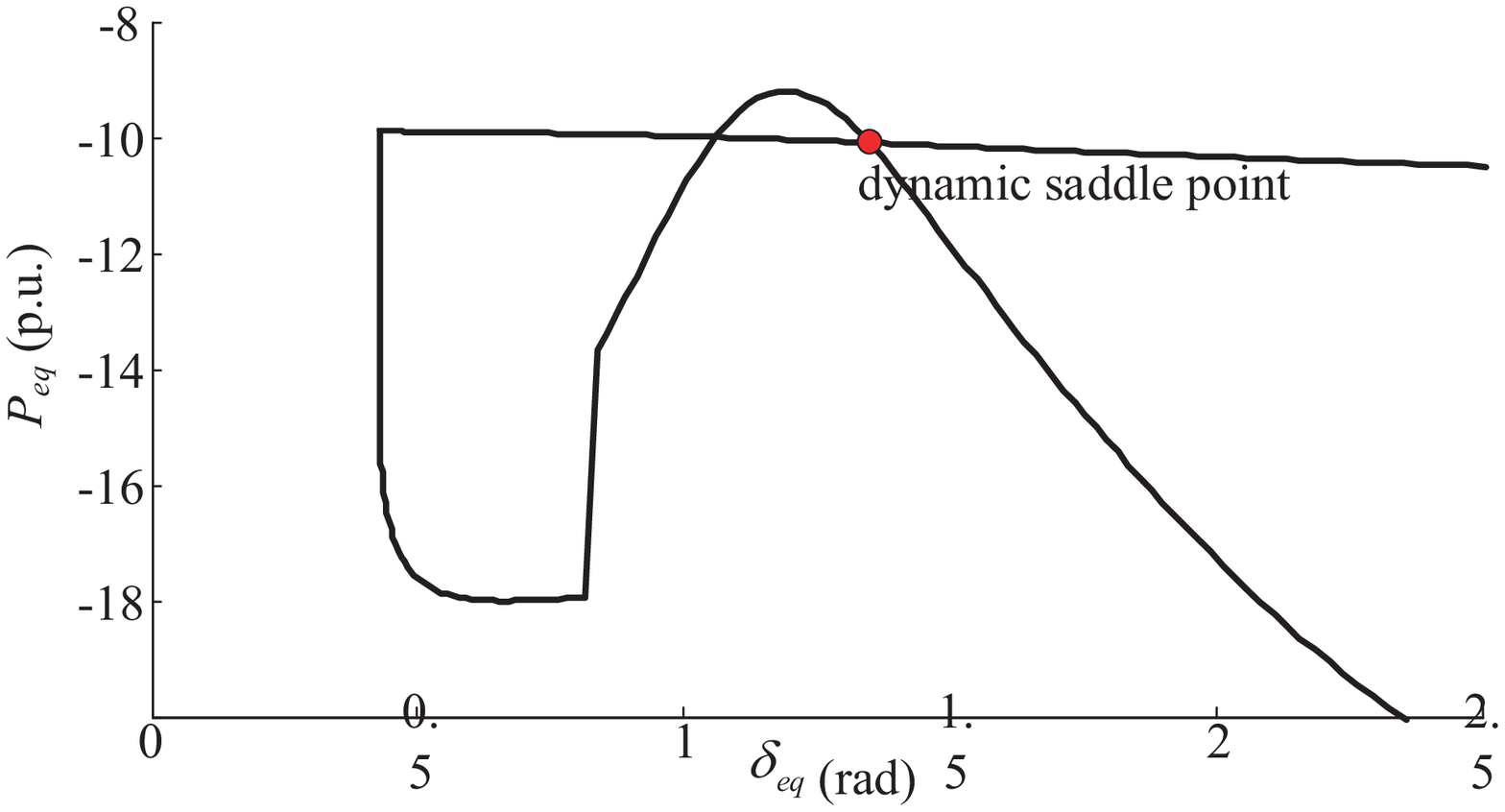}\\
  \setlength{\abovecaptionskip}{-5pt}
  \setlength{\belowcaptionskip}{0pt}
  \vspace{-5pt}
  \caption{Rotor angles of the interconnected system in synchronous reference [TS-3, line-LIAOC\_TANZ, 0.22s]}
  \label{Fig_Kim_Cur_OMIB_16}
\end{figure}

From Fig. \ref{Fig_Kim_Cur_OMIB_16}, the Kimbark curve of the OMIB system goes across dynamic saddle point \cite{Yin2011An} at 0.89s. At this instant the interconnected system is judged to go unstable, the inter-area instability is identified and the stability margin of the interconnected system $\eta_{OMIB}$ is also obtained.

\subsection{Proposed Method (parallel monitoring)}

Using proposed method, the rotor angles of the interconnected system is first depicted in COI reference, as shown in Fig. \ref{Fig_Rot_Ang_IS_COI_17}.

\begin{figure}
\vspace{5pt}
\captionsetup{name=Fig.,font={small},singlelinecheck=off,justification=raggedright}
  \includegraphics[width=3.6in,height=2.4in,keepaspectratio]{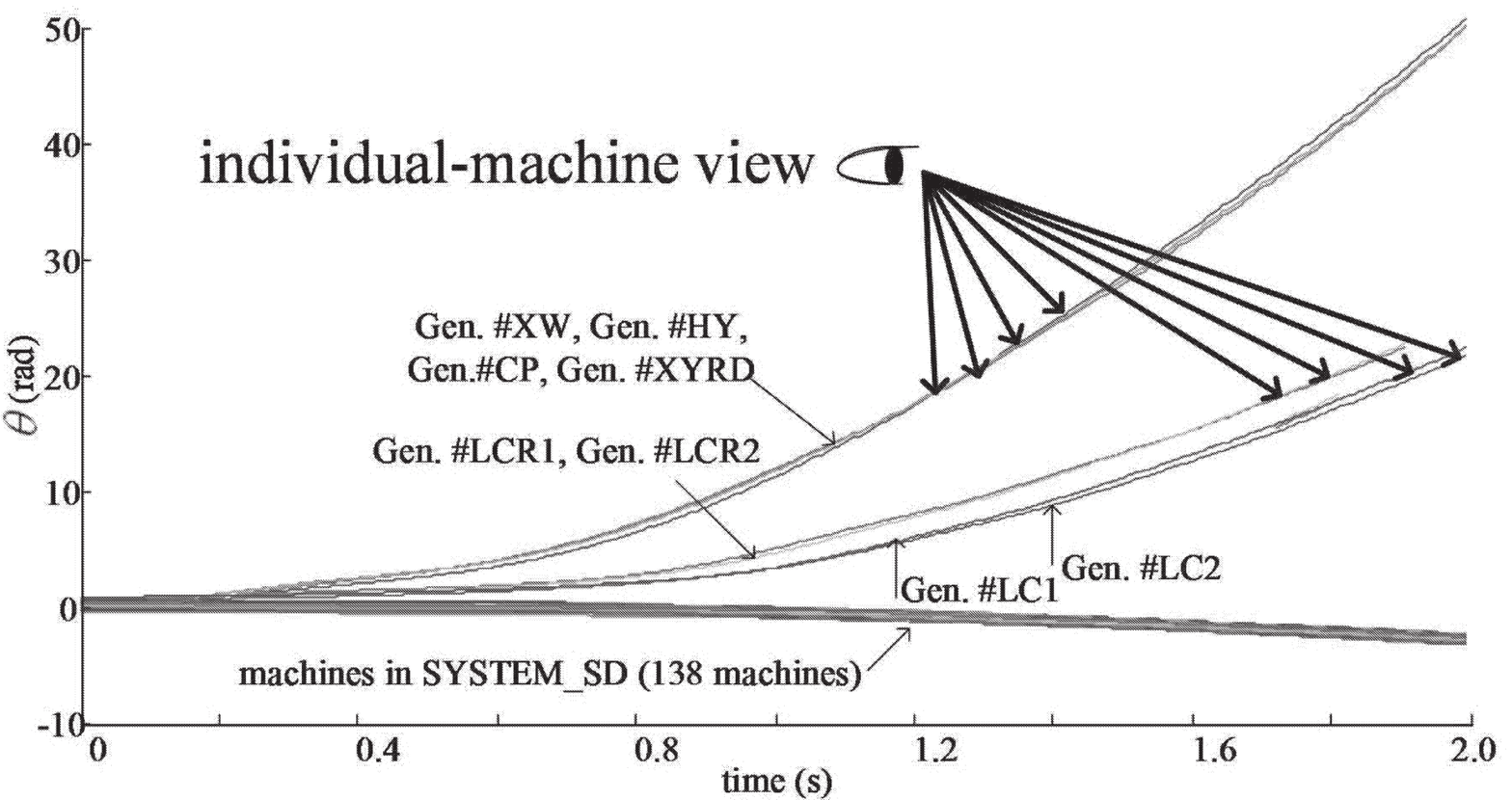}\\
  \setlength{\abovecaptionskip}{-5pt}
  \setlength{\belowcaptionskip}{0pt}
  \vspace{-5pt}
  \caption{Rotor angles of the interconnected system in synchronous reference [TS-3, line-LIAOC\_TANZ, 0.22s]}
  \label{Fig_Rot_Ang_IS_COI_17}
\end{figure}

After fault clearing, all machines in SYSTEM\_LC are identified as critical machines. Unlike the IEEAC method that separates all machines in the interconnected system into two groups, using proposed method the system operator monitors all critical machines in SYSTEM\_LC in parallel, as shown in Fig. \ref{Fig_Rot_Ang_IS_COI_17}. The occurrence of DLPs of these critical machines along time horizon is shown in Fig. \ref{Fig_Ocu_DLP_18}. The Kimbark curves of Gen. \#XYRD and Gen. \#LCR2 are shown in Figs. \ref{Fig_Sim_Kim_Cur_19} (a) and (b), respectively.

\begin{figure}
\vspace{5pt}
\captionsetup{name=Fig.,font={small},singlelinecheck=off,justification=raggedright}
  \includegraphics[width=3.6in,height=2.4in,keepaspectratio]{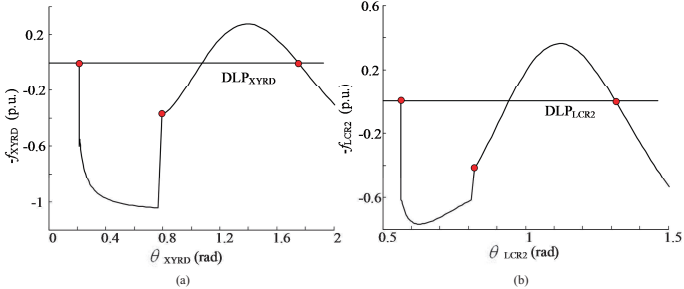}\\
  \setlength{\abovecaptionskip}{-5pt}
  \setlength{\belowcaptionskip}{0pt}
  \vspace{-5pt}
  \caption{ Occurrence of DLPs along time horizon.}
  \label{Fig_Sim_Kim_Cur_19}
\end{figure}

\begin{figure}
\vspace{5pt}
\captionsetup{name=Fig.,font={small},singlelinecheck=off,justification=raggedright}
  \includegraphics[width=3.6in,height=2.4in,keepaspectratio]{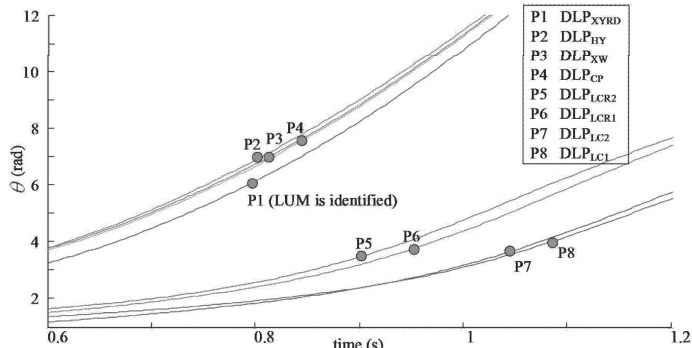}\\
  \setlength{\abovecaptionskip}{-5pt}
  \setlength{\belowcaptionskip}{0pt}
  \vspace{-5pt}
  \caption{Simulated Kimbark curves. (a) Gen. \#XYRD. (b) Gen. \#LCR2.}
  \label{Fig_Ocu_DLP_18}
\end{figure}

According to Fig. \ref{Fig_Ocu_DLP_18}, along time horizon the system operator focuses on following instants.

\emph{$\text{DLP}_{\text{XYRD}}$ occurs (0.79s)}: Gen. \#XYRD is judged as unstable.

\emph{DLPHY-DLPLC2 occur (from 0.80s to 1.03s)}: The corresponding critical machines are judged as unstable consecutively.

\emph{$DLP_{LC1}$ occurs (1.07s)}: Gen. \#LC1 is judged as unstable.

The stability of the system is judged as follows:

\emph{$\text{DLP}_{\text{XYRD}}$ occurs (0.79s)}: $\text{DLP}_{\text{XYRD}}$ is identified as the leading LOSP, and the interconnected system is judged as unstable. Yet, the inter-area instability cannot be identified because the instability of other critical machines in SYSTEM\_LC is still unknown.

\emph{$DLP_{LC1}$  occurs (1.07s)}: The inter-area instability is identified because all critical machines in SYSTEM\_LC are judged as unstable at the instant and $\boldsymbol{\eta}_{sys}$  is finally obtained.

From analysis above, using the proposed method the instability of the interconnected system can be judged earlier than IEEAC method ($\text{DLP}_{\text{XYRD}}$ occurs earlier than dynamic saddle point), yet the \emph{inter-area instability} is identified later than IEEEAC method ($\text{DLP}_{\text{LC1}}$ occurs later than dynamic saddle point). In addition, in IEEAC method, the stability judgement of the system, identification of the inter-area instability and $\eta_{OMIB}$ are obtained simultaneously when the dynamic saddle point occurs, as demonstrated in Fig. \ref{Fig_Kim_Cur_OMIB_16}. Comparatively, using the proposed method the instability of the interconnected system can be identified once the leading LOSP ($\text{DLP}_{\text{XYRD}}$) occurs, but the inter-area instability and $\boldsymbol{\eta}_{sys}$ cannot be obtained until the last DLP (DLPLC1) occurs. This fully proves that the stability judgement of the system is independent of the identification of both the inter-area instability and calculation of $\boldsymbol{\eta}_{sys}$ when using the proposed method.

\subsection{Proposed Method (not-all-critical-machines monitoring) }

From analysis in Section VIII.B, theoretically the inter-area instability can be identified only when the last DLP in SYSTEM\_LC, i.e., $\text{DLP}_{\text{LC1}}$ occurs. However, in real online security control, at the instant of $\text{DLP}_{\text{LCR1}}$ occurs, the system operator can comprehend that SYSTEM\_LC is severely disturbed as most critical machines (six machines) in SYSTEM\_LC already have gone unstable. In this grim situation, the system operator might terminate monitoring the rest of the critical machines and give up identifying inter-area instability. Further, forceful proactive controlling actions may be enforced to SYSTEM\_LC (such as emergency DC power support) as most critical machines in SYSTEM\_LC are identified to go unstable.

\subsection{Separation of a pair of machines}

 From the analysis in the first paper \cite{Wang2017Stability}, an individual machine in COI reference is an IVCS that is formed by a ``pair” of machines in synchronous reference, i.e., the individual machine and the virtual COI machine. In the following analysis, we further demonstrate the mechanisms of IEEAC method and that of the proposed method in synchronous reference. The trajectories of COI machines (Machine-A, Machine-S and the virtual COI machine) in synchronous reference are shown in Fig. \ref{Fig_Sep_Pai_Mac_20}. Notice that, rotor angles of machines in SYSTEM\_SD are not shown in the figure.

\begin{figure}
\vspace{5pt}
\captionsetup{name=Fig.,font={small},singlelinecheck=off,justification=raggedright}
  \includegraphics[width=3.6in,height=2.4in,keepaspectratio]{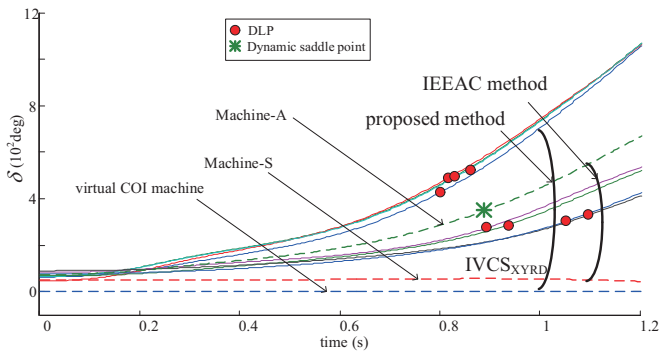}\\
  \setlength{\abovecaptionskip}{-5pt}
  \setlength{\belowcaptionskip}{0pt}
  \vspace{-5pt}
  \caption{ Separation between a pair of machines [TS-3, line-LIAOC\_TANZ, 0.22s].}
  \label{Fig_Sep_Pai_Mac_20}
\end{figure}

It can be seen from Fig. \ref{Fig_Sep_Pai_Mac_20} that both the IEEAC method and the proposed method depict the power system transient stability via the separation of a \emph{pair} of machines. The only difference between them is the formation of the pairs. To be specific, for IEEAC method, the inter-area instability of the interconnected system is depicted as the separation between a ``pair” of equivalent machines, i.e., Machine-A and Machine-S are equivalent machines of the two regional systems. Comparatively, using the proposed method the inter-area instability is depicted as the separation between eight ``pairs” of machines, i.e., the pairs formed by eight individual machines in SYSTEM\_LC and the virtual COI machine. Moreover, using the proposed method the instability (not inter-area instability) of the interconnected system can be directly depicted by the separation of only \emph{one pair of machines}.

From Fig. \ref{Fig_Sep_Pai_Mac_20}, the rotor angles of Machine-S and that of the virtual COI machine are quite close, because non-critical machines in SYSTEM\_SD are majorities after fault clearing. Therefore, since Machine-A is the equivalence of all machines in SYSTEM\_LC, the dynamic saddle point in the OMIB system can also be seen as an ``equivalence” of the DLPs of all critical machines in SYSTEM\_LC, as shown in Fig. \ref{Fig_Sep_Pai_Mac_20}. This can explain the reason why the instability of the interconnected system can be determined earlier while the inter-area instability is identified later than IEEAC method when using the proposed method in TSA.

\section{Conclusion and discussion}

This paper applies the proposed direct-time-domain method for TSA and CCT computation. The stability margin of the system is defined as a vector with its components being the stability margins of all critical machines. Such definition can facilitate parallel monitoring of critical machines in TSA. And this definition further leads to the concept of non-global stability margin, which allows the system operators to give up monitoring some critical machines if they have already comprehend the key transient characteristics of the system. Especially, for the CCT computation, only the MDM is monitored, wherein, the not-all-critical-machines monitoring for MDM is effective to grasp the transition of the system from being stable to being unstable. We also clarify that MDM and LUM might be two different machines for some cases.

Compared with the CUEP method, the proposed method can directly identify MOD via the stability identification of each critical machine of the system. Similar to the IEEAC method, the proposed method also depicts the transient instability of the system through the separation of a pair of machines. The essential difference between the two methods is the formation of the pairs.


\begin{thebibliography}{99}

\bibitem[1]{Wang2017Stability} S. Wang, J. Yu, and W. Zhang, ``“Analytical Mechanism of Partial Energy Function PART I: Equal Area Criterion,”  in {\it arXiv}, pp. 1-8, 2017.

\bibitem[2]{Stanton1989Analysis}S. E. Stanton and W. P. Dykas, ``Analysis of a local transient control action by partial energy functions,”  {\it IEEE Trans. Power Syst.} vol. 4, no. 3, pp. 996-1002, 1989.

\bibitem[3]{Stanton1989Transient}  S. E. Stanton, ``Transient stability monitoring for electric power systems using a partial energy function,”  {\it IEEE Trans. Power Syst.} vol. 4, no. 4, pp. 1389-1396, 1989.


\bibitem[4]{Fouad1989Critical}  A. A. Fouad, V. Vittal, and T. Oh, ``Critical energy for direct transient stability assessment of a multimachine power system,”  {\it IEEE Trans. on Power App. Syst.} vol. 4, no. 4, pp. 1389-1396, 1989.


\bibitem[5]{Yin2011An} M. Yin, C. Y. Chung, K. P. Wong, Y. Xue, Y. Zou, ``An Improved Iterative Method for Assessment of Multi-Swing Transient Stability Limit,”  {\it IEEE Trans. on Power App. Syst.}  vol. 26, no. 4, pp. 2023-2030, 2011.


\end{thebibliography}
\end{document}